\documentclass[%
 reprint,
superscriptaddress,
preprintnumbers,
nofootinbib,
 amsmath,amssymb,
 aps,
 prl,
longbibliography,
twocolumn 
]{revtex4}  

\usepackage[pdftex]{graphicx}
\usepackage[utf8]{inputenc}
\usepackage{dcolumn}
\usepackage{bm}
\usepackage{amsmath}

\usepackage[colorlinks,allcolors=blue,linktocpage]{hyperref}
\usepackage{cleveref}
\crefname{equation}{Eq.}{Eqs.}%

\begin{document}
\title{Radiative stability of tiny cosmological constant from the Swampland and quantized compactification}
\author{Cao H. Nam}
\email{nam.caohoang@phenikaa-uni.edu.vn}  
\affiliation{Phenikaa Institute for Advanced Study and Faculty of Fundamental Sciences, Phenikaa University, Yen Nghia, Ha Dong, Hanoi 12116, Vietnam}
\date{\today}

\begin{abstract}
We address a quantization mechanism that can allow us to understand why the cosmological constant is not large under the quantum corrections from studying the circle compactification solution of the Standard Model coupled to Einstein gravity which is subject to the constraint of the Swampland conjectures. A novel result in the present work compared to the previous investigations in the literature is that the radius of the compactified dimension and the 4D cosmological constant $\Lambda_4$ must in fact be quantized. The quantization rule of the cosmological constant is given by $\Lambda_4\propto n^2$ with $n=1,2,3,4,\ldots$, which means that the values of $\Lambda_4$ are not arbitrary but only its specific values are allowed. In general, the quantum corrections as well as other effects would break this quantization rule. Hence, it could prevent the quantum fluctuations from generating zero-point energy contributions to the cosmological constant.
\end{abstract}

\maketitle

The current cosmological and astrophysical observations \cite{Planck2018} have implied a positive cosmological constant of the value around $\sim10^{-120}M^4_{\text{Pl}}\simeq\left(2.4\times10^{-3}\text{eV}\right)^4$ \cite{Tanabashi2018} (responsible for the late accelerating expansion of the universe) which is a very small number compared to the electroweak scale around $\sim10^2$ GeV and the Planck energy scale $M_{\text{Pl}}\sim10^{19}$ GeV. There is no problem at all with respect to having such a tiny value for the cosmological constant at the classical level. However, the problem arises when the quantum corrections (and other corrections such as non-perturbative effects coming from the QCD instantons) for the cosmological constant are included, which would make the value of the cosmological constant much bigger, and hence its small value is not stable to the large quantum corrections \cite{Zeldovich6768,Birrell1994}. For instance, with the cutoff at the Planck energy scale, the quantum fluctuations would additionally contribute to the cosmological constant with a quantity around $\sim M^4_{\text{Pl}}$. In order to yield the correct physical result, a cancellation between the bare cosmological constant and the quantum corrections must be fine-tuned to an unimaginable precision which is around $\sim120$ orders of magnitude in order. Therefore, the cosmological constant problem has been regarded as one of the most important problems in modern physics \cite{Weinberg1989,Bousso2008}.

The traditional approach to understanding why the cosmological constant is not large under quantum corrections is based on a certain symmetry (either continuous or discrete), which provides a mechanism to cancel the quantum contributions precisely. The most promising symmetry is perhaps supersymmetry which if unbroken would lead to a very precise cancellation between boson and fermions contributing to the vacuum energy since their contributions are opposite signs. Unfortunately, the superpartners of the Standard Model (SM) particles have not been observed so far since supersymmetry must be spontaneously broken down at least at the TeV scale. It was also suggested that the one-loop contribution to the cosmological constant might vanish in certain string models with broken supersymmetry \cite{Kachru1998,Kachru1999}. However, it was argued that the higher-order quantum corrections would spoil this vanishing \cite{Harvey1999,Iengo2000}. In addition, these string models need to have the supersymmetric non-Abelian sector \cite{Blumenhagen,Angelantonj1999,Angelantonj2004}. Another promising symmetry to solve the cosmological constant problem is scale invariance or conformal symmetry which forbids the quantities or the terms in Lagrangian with the length scale dimension and hence the cosmological constant has to be zero. However, the observed world simply is not invariant under the scale transformation. For other ideas which have been proposed in the literature, see \cite{Nobbenhuis2006,Burgess2013} for reviews. Although there have been attempts which have been made to solve the cosmological constant problem there is indeed no known solution to this problem at present. 

The Swampland program aims \cite{Vafa2005,Ooguri2007,Palti2019} to seek the universal features of quantum gravity and the consistency criteria which distinguish effective field theories UV-completed consistent with quantum gravity from those which cannot or are in the Swampland. In particular, this program leads to interesting implications for inflationary cosmology (corresponding to the early accelerating expansion of the universe) which is constrained by the Swampland criterion for de Sitter \cite{Obied2018,Garg2019} and trans-Planckian censorship conjecture \cite{Bedroya2020,Brandenberger2021}. In this paper, based on the Swampland program we will point to a novel and simple mechanism which could forbid quantum fluctuations (as well as other effects) to generate zero-point energy contributions to the cosmological constant. Although this mechanism is far from a solution to the cosmological constant problem, we hope that it could give a step toward a complete solution.

If a theory satisfies the Swampland constraints and is completed into quantum gravity in UV, then a lower-dimensional theory that is obtained from its compactification to lower dimensions must do as well. On the contrary, if the theory upon compactification is inconsistent with quantum gravity then its parent higher dimensional theory itself is pathological. Because the SM coupled to General Relativity (GR) is a good effective low energy theory and thus cannot be in the Swampland, its compactification solution down to lower dimensions \cite{Arkani-Hamed2007,Arnold2010,Fornal2011} must be consistent with quantum gravity. In this sense, the Swampland conjectures impose the constraints not only on the SM coupled to GR itself but also on its compactification solution. This has motivated the intensive investigations in the literature with the significant implications between the cosmological constant and the neutrino masses \cite{Martin-Lozano2017,Valenzuela2017,Hamada2017,Gonzalo2018a,Gonzalo2018b} arising when applying the non-SUSY AdS conjecture \cite{Ooguri2017} (motivated by the weak gravity conjecture \cite{Arkani-Hamed2007b}) and the AdS distance conjecture \cite{Lust2019} (realized in string theory and also supported by the evidence of bottom-up physics \cite{Rudelius2021,Gonzalo2021,Nam2022}). 

In the present work, in this direction, we study the compactification of the SM coupled to GR on a circle $S^1$. However, a new point here is that we consider the compactified coordinate dependence of the 3D tensor component of the 4D metric in a natural way which is usually ignored in the literature. Interestingly, from investigating the wavefunction profile of the 3D tensor component along the compactified dimension we find a novel aspect that the radius of the compactified dimension and the 4D cosmological constant are in fact quantized. Here, we obtain a quantization rule of the 4D cosmological constant $\Lambda_4$ as $\Lambda_4\propto n^2$ with $n=1,2,3,4,\ldots$. In this way, the values of the 4D cosmological constant are not arbitrary but only its specific values are allowed. This quantization or this discrete spectrum could prevent the quantum fluctuations from generating zero-point energy contributions to the 4D cosmological constant because in general, the additional contribution from the quantum fluctuations would not respect the quantization rule. This is in analogy to that the hydrogen atom would be unable to absorb the energy which is not equal to the energy difference between two initial and final states due to the quantization of its energy levels.

The action of the SM coupled to GR in the presence of the observed cosmological constant $\Lambda_4$ is given as
\begin{eqnarray}
S&=&\int d^4x\sqrt{-g}\left[\frac{M^2_{\text{Pl}}}{2}\left(\mathcal{R}^{(4)}-2\Lambda_4\right)+\mathcal{L}_{SM}\right],\label{SMEHLa-act}
\end{eqnarray}
where $\mathcal{R}^{(4)}$ and $\mathcal{L}_{SM}$ refer to the scalar curvature of 4D spacetime and Lagrangian of the SM, respectively. We consider the above system compactified on a circle $S^1$ where one of three spatial dimensions of 4D spacetime is periodic with the period of $2\pi$, which means that the points $x^3$ and $x^3+2\pi$ are identified. The 4D metric is generally parametrized in terms of the 3D fields as follows
\begin{eqnarray}
ds^2_4=g_{ij}dx^i dx^i+R^2\left[dx^3+g_{_A} A_idx^i\right]^2,\label{KKmetric}
\end{eqnarray}
where $g_{ij}$ is the 3D metric component, $A_i$ the graviphoton, $R$ is the radion field which determines the radius of $S^1$, $g_{_A}$ is a coupling constant, and $i,j=0,1,2$.\footnote{We use the metric sign as $(-,+,+,+)$.} Using this ansatz of the 4D metric, we can expand explicitly the 4D scalar curvature $\mathcal{R}^{(4)}$ given in the action (\ref{SMEHLa-act}) as
\begin{eqnarray}
\mathcal{R}^{(4)}&=&\mathcal{R}+\frac{1}{4R^2}\left(\partial_3g^{ij}\partial_3 g_{ij}+g^{ij}g^{kl}\partial_3 g_{ij}\partial_3 g_{kl}\right)\nonumber\\
&&-\frac{g^2_{_A}R^2}{4}F_{ij}F^{ij},\label{curexp}
\end{eqnarray}
where $\mathcal{R}\equiv g^{ij}\left(\hat{\partial}_k\hat{\Gamma}^k_{ji}-\hat{\partial}_j\hat{\Gamma}^k_{ki}+\hat{\Gamma}^k_{ji}\hat{\Gamma}^l_{lk}-\hat{\Gamma}^k_{li}\hat{\Gamma}^l_{jk}\right)$ with $\hat{\Gamma}^k_{ij}\equiv\frac{g^{kl}}{2}(\hat{\partial}_i g_{jl}+\hat{\partial}_j g_{il}-\hat{\partial}_l g_{ij})$, $\hat{\partial}_i\equiv\partial_i-g_{_A} A_i\partial_3$, and $F_{ij}=\partial_i A_j-\partial_j A_i$. It is important to emphasize here that the previous investigations about the compactification of the SM coupled with GR on $S^1$ \cite{Arkani-Hamed2007,Arnold2010,Fornal2011,Martin-Lozano2017,Valenzuela2017,Hamada2017,Gonzalo2018a,Gonzalo2018b} have not considered the compactified dimension dependence of the 3D metric $ g_{ij}$ and hence the second term in the right-hand side of Eq. (\ref{curexp}) was absent. But, in the present work, we will point out that the presence of this term would lead to the quantization of the 4D cosmological constant which could provide a mechanism to prevent the vacuum fluctuations as well as other effects contributing additionally to the cosmological constant.

Let us find the wavefunction profile of the 3D metric $g_{ij}$. In order to do this, first we obtain the equation of motion for $g_{ij}$ in the vacuum $R=\text{constant}$, $A_i=0$, and the vanishing matter fields by varying the action (\ref{SMEHLa-act}) in the 3D tensor component of the 4D metric as follows
\begin{eqnarray}
\bar{\mathcal{R}}_{ij}-\frac{1}{2}g_{ij}\bar{\mathcal{R}}+\Lambda_4g_{ij}+\frac{1}{4R^2}\left[g_{ik}g_{jl}\partial^2_3g^{kl}\right.&&\nonumber\\
\left.-\partial^2_3g_{ij}+g^{kl}\partial_3 g_{kl}\partial_3 g_{ij}+2\partial_3\left(g_{ij}g^{kl}\partial_3g_{kl}\right)\right.&&\nonumber\\
\left.-\frac{1}{2}g_{ij}\left\{\partial_3g^{kl}\partial_3g_{kl}-\left(g^{kl}\partial_3g_{kl}\right)^2\right\}\right]=0,&&\label{3Dtensor-equ}
\end{eqnarray}
where $\bar{\mathcal{R}}_{ij}\equiv(\partial_k\bar{\Gamma}^k_{ji}-\partial_j\bar{\Gamma}^k_{ki}+\bar{\Gamma}^l_{ji}\bar{\Gamma}^k_{kl}-\bar{\Gamma}^l_{ki}\bar{\Gamma}^k_{jl})$ with $\bar{\Gamma}^k_{ij}\equiv\frac{g^{kl}}{2}(\partial_i g_{lj}+\partial_j g_{ki}-\partial_l g_{ij})$ and $\bar{\mathcal{R}}\equiv g^{ij}\bar{\mathcal{R}}_{ij}$. One can solve Eq. (\ref{3Dtensor-equ}) by separating the variables as $g_{ij}(x^i,x^3)=\chi(x^3)g^{(3)}_{ij}(x^i)$ where $g^{(3)}_{ij}(x^i)$ is identified as the metric of the 3D effective theory and $\chi(x^3)$ is its wavefunction profile along the compactified dimension. Then, substituting this variable separation into Eq. (\ref{3Dtensor-equ}), we obtain the 3D Einstein field equations determining the 3D effective geometry of spacetime and the equation for the wavefunction profile of the 3D metric as follows
\begin{eqnarray}
\mathcal{R}^{(3)}_{ij}-\frac{1}{2}g^{(3)}_{ij}\mathcal{R}^{(3)}+\Lambda_3g^{(3)}_{ij}&=&0,\label{effEinsEq}\\
\chi''+\frac{11}{4}\frac{\chi'^2}{\chi}+\kappa^2\chi&=&\frac{\Lambda_3}{R^{-2}},\label{chiEq}
\end{eqnarray}
where $\mathcal{R}^{(3)}_{ij}$ ($\mathcal{R}^{(3)}$) is the Ricci (scalar) curvature of 3D effective spacetime, $\kappa\equiv\sqrt{\Lambda_4}/R^{-1}$, and $\Lambda_3$ is a constant which characterizes the dynamics of the 3D metric $g_{ij}$ along the compactified dimension and from Eq. (\ref{effEinsEq}) we see that it plays the role of a 3D cosmological constant. 

It is difficult to find an analytical solution for $\chi(x^3)$ in the general case. However, in the situation of $\Lambda_3/R^{-2}\ll1$ (which can be seen from Table. 1 in Ref. \cite{Nam2023}), we can perturbatively solve Eq. (\ref{chiEq}) in the order of $\Lambda_3/R^{-2}$. The solution of $\chi(\theta)$ at the leading order is easily found as follows
\begin{eqnarray}
\chi(x^3)=\left[\frac{1}{2}\left\{1+\cos\left(\sqrt{15}\kappa x^3\right)\right\}\right]^{2/15}.\label{3Dmetr-profile}
\end{eqnarray}
Due to the topology of $S^1$, the wavefunction profile $\chi(x^3)$ must be periodic with the period of $2\pi$, i.e. $\chi(x^3)=\chi(x^3+2\pi)$, which implies $\sqrt{15}\kappa=n$ with $n=1,2,3,\ldots.$ This thus leads to the quantization of the radius $R$ of the compactified dimension and 4D cosmological constant $\Lambda_4$ as
\begin{eqnarray}
\Lambda_4R^2=\frac{n^2}{15}.\label{radquan}
\end{eqnarray}
Equation (\ref{radquan}) means that the radius of the compactified dimension and the 4D cosmological constant are not arbitrary but must obtain the discrete values according to the quantization relation (\ref{radquan}). On the other hand, the radius of the compactified dimension and the 4D cosmological constant have no continuous spectrum but a discrete one. This is a novel point of the circle compactification solution of the SM coupled to GR which is pointed to the first time. 

The quantization relation (\ref{radquan}) results from the nontrivial dynamics of the 3D metric along the compact direction in the presence of the cosmological constant. From Eq. (\ref{chiEq}) that describes the dynamic propagation of the 3D metric along the compact direction, we see that the 4D cosmological constant $\Lambda_4$ plays the role of a force proportional linearly to the profile $\chi$ because its presence curves the spacetime geometry, whereas the inverse radius $R^{-1}$ of the compactified dimension plays the role of a mass for the 3D metric because its propagation is confined along the compact direction. (Note that, the term $11\chi'^2/(4\chi)$ in Eq. (\ref{chiEq}) comes from the nonlinear property of the gravitational field which means that gravity is itself a source creating the curvature of the spacetime geometry.) As a result, it leads to the oscillation behavior of the 3D metric along the compactified dimension with the corresponding period depending on the ratio of $\sqrt{\Lambda_4}$ to $R^{-1}$. In addition, the dynamic propagation of the 3D metric is constrained by the periodic property of the compactified dimension. Hence, both of these aspects result in the quantization of the radius $R$ of the compactified dimension and 4D cosmological constant $\Lambda_4$.

In fact, the size of the compactified dimension can be physically determined by the 3D effective potential of the radion field which is generated by the dynamics of $3$D tensor component along the compactified dimension and the one-loop quantum corrections with the Casimir energy density calculated in Ref. \cite{Arkani-Hamed2007} (see Appendix D for detailed computation). Taking into account the stabilization mechanism and from Eq. (\ref{radquan}), we can find a general quantization rule of the 4D cosmological constant in terms of the mass of the light particles in the spectrum and the parameter $\Lambda_3$ as follows
\begin{eqnarray}
\Lambda_4=f(m_i,n_i,\Lambda_3)n^2\propto n^2,\label{ccuqan2}
\end{eqnarray}
where $f(m_i,n_i,\Lambda_3)$ is a function of $m_i$ (the mass of the $i$th light particle), $n_i$ (the number of degrees of freedom corresponding to the $i$th light particle), and $\Lambda_3$. Note that, the function $f(m_i,n_i,\Lambda_3)$ can be completely determined from the minimum of the radion potential. The quantization rule (\ref{ccuqan2}) implies that the ratio of the 4D cosmological constant to a specific combination of the 3D cosmological constant and the masses of the particles contributing to the radion potential must lead to 1, 4, 9, 16, 25, $\ldots$. In general, under the quantum corrections this ratio would not lead to 1, 4, 9, 16, 25, $\ldots$ which are only the allowed values. Hence, we expect that the quantization rule (\ref{ccuqan2}) can provide a mechanism to prevent the quantum fluctuations from generating additional contributions to the 4D cosmological constant. 

It is important to emphasize that Eq. (\ref{radquan}) can be realized as a quantum gravity sign that arises at the low-energy regime. Hence, one expects that the quantization rule for the radius of the compactified dimension and the cosmological constant could provide a step toward solving not only the cosmological constant problem but also other important problems (which require a complete theory of quantum gravity) such as a microscopic description of the black hole entropy. By considering the compactification of $\text{dS}_5$ on a circle $S^1$ to obtain 4D effective field theory, we can find a quantization relation for the size of the extra dimension as $R^{-1}\propto\sqrt{\Lambda_5}/n$ where $\Lambda_5$ is the bulk cosmological constant and $n=1,2,3,\ldots$, in analogy to what we have done for the $S^1$ compactification of the SM. Each value of the positive integer $n$ corresponding to a possible value of the size of the extra dimension implies a possible configuration of 4D effective field theory. Hence, we can calculate the black hole entropy by summing over all relevant configurations. We leave an exploration of this proposal to future work. 
\section*{Appendix A: The expansion of $\mathcal{R}^{(4)}$ in terms of 3D components}

Because Eq. (\ref{curexp}) is the main technical starting point, in this appendix we present detailed computations of how to obtain it. It is easy to do the calculations in the frame $\{\hat{\partial}_i,\hat{\partial}_3\}\equiv\{\hat{\partial}_\mu\}$ where $\hat{\partial}_i\equiv\partial_i-g_{_A} A_i\partial_3$ and $\hat{\partial}_3\equiv\partial_3$ which transform as a 3D vector and a 1D vector, respectively. The corresponding coframe is given by $\{dx^i,dx^3+g_{_A}A_i dx^i\}$ which dual to $\{\hat{\partial}_i,\hat{\partial}_3\}$, respectively. The components of 4D metric and its dual are given in these bases as
\begin{eqnarray}
  \hat{g}_{\mu\nu}&=&\textrm{diag}\left(g_{ij},R^2\right),\nonumber\\
  \hat{g}^{\mu\nu}&=&\textrm{diag}\left(g^{ij},R^{-2}\right).
\end{eqnarray}
Then, we can determine the Christoffel connection $\hat{\Gamma}^\rho_{\mu\nu}$ and the Riemann curvature tensor $\hat{\mathcal{R}}^\lambda_{\mu\rho\nu}$ in these bases as follows
\begin{eqnarray}
\hat{\Gamma}^\rho_{\mu\nu}&=&\frac{\hat{g}^{\rho\lambda}}{2}\left(\hat{\partial}_\mu\hat{g}_{\nu\lambda}+\hat{\partial}_\nu\hat{g}_{\mu\lambda}-\hat{\partial}_
\lambda\hat{g}_{\mu\nu}\right)\nonumber\\
&&+\frac{\hat{g}^{\rho\lambda}}{2}\left(C^\sigma_{\lambda\mu}\hat{g}_{\nu\sigma}+C^\sigma_{\lambda\nu}\hat{g}_{\mu\sigma}\right)+\frac{C^\rho_{\mu\nu}}{2},\nonumber\\
\hat{\mathcal{R}}^\lambda_{\mu\rho\nu}&=&\hat{\partial}_\rho\hat{\Gamma}^\lambda_{\nu\mu}-\hat{\partial}_\nu\hat{\Gamma}^\lambda_{\rho\mu}+
\hat{\Gamma}^\sigma_{\nu\mu}\hat{\Gamma}^\lambda_{\rho\sigma}-\hat{\Gamma}^\sigma_{\rho\mu}\hat{\Gamma}^\lambda_{\nu\sigma}\nonumber\\
&&-C^\sigma_{\rho\nu}\hat{\Gamma}^\lambda_{\sigma\mu},
\end{eqnarray}
where $C^\rho_{\mu\nu}$ defines the commutation of two frame fields as
\begin{equation}
\left[\hat{\partial}_\mu,\hat{\partial}_\nu\right]=C^\rho_{\mu\nu}\hat{\partial}_\rho.
\end{equation}

The 4D scalar curvature $\mathcal{R}^{(4)}$ reads
\begin{eqnarray}
\mathcal{R}^{(4)}=\hat{g}^{\mu\nu}\hat{\mathcal{R}}^\rho_{\mu\rho\nu}=g^{ij}\hat{\mathcal{R}}^\rho_{i\rho j}+g^{33}\hat{\mathcal{R}}^\rho_{3\rho3}.
\end{eqnarray}
The first and second terms are explicitly expanded as follows
\begin{eqnarray}
g^{ij}\hat{\mathcal{R}}^\rho_{i\rho j}&=&g^{ij}\left(\hat{\partial}_\rho\hat{\Gamma}^\rho_{ji}-\hat{\partial}_j\hat{\Gamma}^\rho_{\rho i}+\hat{\Gamma}^\rho_{ji}\hat{\Gamma}^\lambda_{\lambda\rho}-\hat{\Gamma}^\rho_{\lambda i}\hat{\Gamma}^\lambda_{j\rho}\right.\nonumber\\
&&\left.-C^\rho_{\lambda j}\hat{\Gamma}^\lambda_{\rho i}\right)\nonumber\\
&=&g^{ij}\left[\left(\hat{\partial}_k\hat{\Gamma}^k_{ji}-\hat{\partial}_j\hat{\Gamma}^k_{ki}+\hat{\Gamma}^k_{ji}\hat{\Gamma}^l_{lk}-\hat{\Gamma}^k_{li}\hat{\Gamma}^l_{jk}\right)\right.\nonumber\\
&&+\left.\left(\hat{\partial}_3\hat{\Gamma}^3_{ji}+\hat{\Gamma}^3_{ji}\hat{\Gamma}^k_{k3}+\hat{\Gamma}^k_{ji}\hat{\Gamma}^3_{3k}+\hat{\Gamma}^3_{ji}\hat{\Gamma}^3_{33}\right)\right.\nonumber\\
&&-\left.\left(\hat{\partial}_j\hat{\Gamma}^3_{3i}+\hat{\Gamma}^k_{3i}\hat{\Gamma}^3_{jk}+\hat{\Gamma}^3_{ki}\hat{\Gamma}^k_{j3}+\hat{\Gamma}^3_{3i}\hat{\Gamma}^3_{j3}\right)\nonumber\right.\\ 
&&-\left.\left(C^3_{kj}\hat{\Gamma}^k_{3i}+C^3_{3j}\hat{\Gamma}^3_{3i}\right)\right],\label{sccurI}\\
g^{33}\hat{\mathcal{R}}^\rho_{3\rho3}&=&g^{33}\left(\hat{\partial}_\rho\hat{\Gamma}^\rho_{33}-\hat{\partial}_3\hat{\Gamma}^\rho_{\rho3}+\hat{\Gamma}^\rho_{33}\hat{\Gamma}^\lambda_{\lambda\rho}-\hat{\Gamma}^\rho_{\lambda3}\hat{\Gamma}^\lambda_{3\rho}\right.\nonumber\\
&&\left.-C^\rho_{\lambda3}\hat{\Gamma}^\lambda_{\rho3}\right)\nonumber\\
&=&g^{33}\left(\hat{\partial}_i\hat{\Gamma}^i_{33}-\hat{\partial}_3\hat{\Gamma}^i_{i3}+\hat{\Gamma}^i_{33}\hat{\Gamma}^j_{ji}+\hat{\Gamma}^3_{33}\hat{\Gamma}^i_{i3}\right.\nonumber\\
&&\left.-\hat{\Gamma}^j_{i3}\hat{\Gamma}^i_{3j}-\hat{\Gamma}^3_{i3}\hat{\Gamma}^i_{33}-C^3_{i3}\hat{\Gamma}^i_{33}\right).\label{sccurII}   
\end{eqnarray}
In (\ref{sccurI}) and (\ref{sccurII}), we find the following combinations
\begin{eqnarray}
&&g^{ij}\left(\hat{\partial}_3\hat{\Gamma}^3_{ji}+\hat{\Gamma}^3_{ji}\hat{\Gamma}^k_{k3}+\hat{\Gamma}^3_{ji}\hat{\Gamma}^3_{33}\right)\equiv\nabla_\mu Y^\mu_1\nonumber \\
&&+\frac{g^{33}}{2}\hat{\partial}_3g^{ij}\hat{\partial}_3g_{ij},
\end{eqnarray}
\begin{eqnarray}
g^{ij}\left(\hat{\partial}_j\hat{\Gamma}^3_{3i}-\hat{\Gamma}^k_{ji}\hat{\Gamma}^3_{3k}\right)-g^{33}\hat{\Gamma}^i_{33}\hat{\Gamma}^3_{3i}&\equiv&\nabla_\mu Y^\mu_2,
\end{eqnarray}
\begin{eqnarray}
&&g^{33}\left(\hat{\partial}_i\hat{\Gamma}^i_{33}+\hat{\Gamma}^i_{33}\hat{\Gamma}^j_{ji}+\hat{\Gamma}^i_{33}\hat{\Gamma}^3_{3i}\right)\equiv\nabla_\mu Y^\mu_3\nonumber\\
&&+\frac{g^{ij}}{2}\hat{\partial}_ig^{33}\hat{\partial}_jg_{33}+g_{_A}g^{ij}\hat{\partial}_3A_ig^{33}\hat{\partial}_jg_{33},
\end{eqnarray}
\begin{eqnarray}
g^{33}\left(\hat{\partial}_3\hat{\Gamma}^i_{i3}-\hat{\Gamma}^3_{33}\hat{\Gamma}^i_{i3}\right)-g^{ij}\hat{\Gamma}^3_{ij}\hat{\Gamma}^k_{k3}&\equiv&\nabla_\mu Y^\mu_4,
\end{eqnarray}
where
\begin{eqnarray}
Y^\mu_1&\equiv&\left(0,g^{ij}\hat{\Gamma}^3_{ji}\right),\nonumber\\
Y^\mu_2&\equiv&\left(g^{ij}\hat{\Gamma}^3_{3j},0\right),\nonumber\\
Y^\mu_2&\equiv&\left(g^{33}\hat{\Gamma}^i_{33},0\right),\nonumber\\
Y^\mu_4&\equiv&\left(0,g^{33}\hat{\Gamma}^i_{i3}\right).
\end{eqnarray}
The terms $\nabla_\mu Y^\mu_a$ with $a=1,2,3,4$ are divergences corresponding to the boundary terms and as a result, they vanish at infinity. Then, $\mathcal{R}^{(4)}$ is a sum of the remaining terms as
\begin{eqnarray}
\mathcal{R}^{(4)}&=&\mathcal{R}-g^{ij}\left(\hat{\Gamma}^k_{3i}\hat{\Gamma}^3_{jk}+\hat{\Gamma}^3_{ki}\hat{\Gamma}^k_{j3}+\hat{\Gamma}^3_{3i}\hat{\Gamma}^3_{j3}+\hat{\Gamma}^3_{ij}\hat{\Gamma}^k_{k3}\right.\nonumber\\ 
&&\left.+C^3_{kj}\hat{\Gamma}^k_{3i}+C^3_{3j}\hat{\Gamma}^3_{3i}\right)-g^{33}\left(\hat{\Gamma}^j_{i3}\hat{\Gamma}^i_{3j}+\hat{\Gamma}^3_{i3}\hat{\Gamma}^i_{33}\right.\nonumber\\
&&\left.+2\hat{\Gamma}^3_{3i}\hat{\Gamma}^i_{33}+C^3_{i3}\hat{\Gamma}^i_{33}\right)+\frac{g^{33}}{2}\hat{\partial}_3g^{ij}\hat{\partial}_3g_{ij}\nonumber\\
&&+\frac{g^{ij}}{2}\hat{\partial}_ig^{33}\hat{\partial}_jg_{33}+g_{_A}g^{ij}\hat{\partial}_3A_ig^{33}\hat{\partial}_jg_{33}\nonumber\\
&=&\mathcal{R}+\frac{1}{4R^2}\left(\partial_3g^{ij}\partial_3 g_{ij}+g^{ij}g^{kl}\partial_3 g_{ij}\partial_3 g_{kl}\right)\nonumber\\
&&-\frac{g^2_{_A}R^2}{4}F_{ij}F^{ij},\label{expR4}
\end{eqnarray}
where $\mathcal{R}\equiv g^{ij}\left(\hat{\partial}_k\hat{\Gamma}^k_{ji}-\hat{\partial}_j\hat{\Gamma}^k_{ki}+\hat{\Gamma}^k_{ji}\hat{\Gamma}^l_{lk}-\hat{\Gamma}^k_{li}\hat{\Gamma}^l_{jk}\right)\equiv g^{ij}R_{ij}$ and $F_{ij}=\partial_i A_j-\partial_j A_i$. Note that, in the calculation of the scalar curvature $\mathcal{R}^{(4)}$, we have considered the components $g_{ij}$, $A_i$, and $R$ to be the general functions of $(x^i,x^3)$. However, we observe that first the terms relating to $\hat{\partial}_3A_i$ automatically cancel together and hence they do not contribute to $\mathcal{R}^{(4)}$. This is because the spacetime curvature which is caused by the non-triviality of the $U(1)$ principal bundle is measured by the curvature of the connection $A_i$, i.e. $F_{ij}$. Second, the first-order derivative terms of $R$ associated with its kinetic terms do not appear in $\mathcal{R}^{(4)}$ because the curvature of the $S^1$ fiber is zero. But, the kinetic terms of $R$ would appear in Einstein frame derived by rescaling the 3D metric as $g_{ij}\rightarrow\Omega^{-2}g_{ij}$ with $\Omega=R/r$ where $r$ is introduced to keep the rescaled 3D metric dimensionless and would be fixed equal to the vacuum expectation value of the radion field.

\section*{Appendix B: Equations of motion for the 3D components of the bulk metric}
First, let us obtain equations of motion for the 3D tensor component $g_{ij}$ from the following variation
\begin{eqnarray}
\delta_{g_3}S_{\text{EH}}&=&\delta_{g_3}\int d^4x\sqrt{-g}\frac{M^2_{\text{Pl}}}{2}\left(\mathcal{R}^{(4)}-2\Lambda_4\right)\nonumber\\
&=&\frac{M^2_{\text{Pl}}}{2}\int d^4x\left[\left(\mathcal{R}^{(4)}-2\Lambda_4\right)\delta_{g_3}\sqrt{-g}\right.\nonumber\\
    &&\left.+\sqrt{-g}\delta_{g_3}\mathcal{R}^{(4)}\right]=0,
\end{eqnarray}
where $\delta_{g_3}$ refers to the variation in terms of $g_{ij}$ and the expansion of $\mathcal{R}^{(4)}$ is given in Eq. (\ref{expR4}). We find
$\delta_{g_3}\sqrt{-g}=-\frac{1}{2}\sqrt{-g}g_{ij}\delta g^{ij}$ and
\begin{eqnarray}
    \delta_{g_3}\mathcal{R}^{(4)}&=&\delta g^{ij}R_{ij}+g^{ij}\delta_{g_3}R_{ij}\nonumber\\
    &&+\frac{1}{4}\delta_{g_3}\left(\partial_3g^{ij}\partial^3 g_{ij}+g^{ij}g^{kl}\partial_3 g_{ij}\partial^3 g_{kl}\right)\nonumber\\
    &&-\frac{g^2_{_A}R^2}{4}\delta_{g_3}\left(F_{ij}F^{ij}\right).\label{varRiccten}
\end{eqnarray}
The second variation term in (\ref{varRiccten}) is calculated as
\begin{eqnarray}
g^{ij}\delta_{g_3}R_{ij}&=&\nabla_k\left(g_{ij}\nabla^k\delta g^{ij}-\nabla_l\delta g^{kl}\right)\nonumber\\    
&=&\nabla_\mu X^\mu_1-\hat{\Gamma}^3_{3k}\left(g_{ij}\nabla^k\delta g^{ij}-\nabla_l\delta g^{kl}\right)\nonumber\\
&=&\nabla_\mu X^\mu_1+\nabla_\mu X^\mu_2+\nabla_\mu X^\mu_3+\hat{\Gamma}^3_{3k}\hat{\Gamma}^3_{3l}g^{kl}g_{ij}\delta g^{ij}\nonumber\\
&&+\nabla_k\left(g^{kl}\hat{\Gamma}^3_{3l}g_{ij}\right)\delta g^{ij}-\hat{\Gamma}^3_{3i}\hat{\Gamma}^3_{3j}\delta g^{ij}\nonumber\\
&&-\nabla_i\left(\hat{\Gamma}^3_{3j}\right)\delta g^{ij},
\end{eqnarray}
where
\begin{eqnarray}
    X^\mu_1&=&\left(g_{ij}\nabla^k\delta g^{ij}-\nabla_l\delta g^{kl},0\right),\nonumber\\
    X^\mu_2&=&\left(-g^{kl}\hat{\Gamma}^3_{3l}g_{ij}\delta g^{ij},0\right),\\
    X^\mu_3&=&\left(\hat{\Gamma}^3_{3l}\delta g^{lk},0\right).\nonumber
\end{eqnarray}
The variation of the third term in (\ref{varRiccten}) reads
\begin{eqnarray}
&&\delta_{g_3}\left(\partial_3g^{ij}\partial^3 g_{ij}+g^{ij}g^{kl}\partial_3 g_{ij}\partial^3 g_{kl}\right)=\nabla_\mu X^\mu_4+\nabla_\mu X^\mu_5\nonumber\\
&&+g_{ik}g_{jl}\partial_3\partial^3g^{kl}-\partial_3\partial^3g_{ij}+g^{kl}\partial_3g_{kl}\partial^3g_{ij}\delta g^{ij}\nonumber\\
&&+2\partial_3\left(g_{ij}g^{kl}\partial^3g_{kl}\right)+\left(g^{kl}\partial_3g_{kl}\right)^2g_{ij}\delta g^{ij} \nonumber\\
&&-2\hat{\Gamma}^3_{33}\left(\partial^3g_{ij}+g_{ij}g_{kl}\partial^3g^{kl}\right)\delta g^{ij},
\end{eqnarray}
where
\begin{eqnarray}
    X^\mu_4&=&\left(0,2\partial^3g_{ij}\delta g^{ij}\right),\nonumber\\
    X^\mu_5&=&\left(0,2g_{ij}g_{kl}\partial^3g^{kl}\delta g^{ij}\right).
\end{eqnarray}
The last variation term is
\begin{eqnarray}
\delta_{g_3}\left(F_{ij}F^{ij}\right)=2{F_i}^kF_{jk}.   
\end{eqnarray}
Finally, we find the equations of motion for $g_{ij}$ as follows
\begin{eqnarray}
\mathcal{R}_{ij}-\frac{1}{2}g_{ij}\mathcal{R}+\Lambda_4g_{ij}+\frac{1}{4R^2}\left[g_{ik}g_{jl}\partial^2_3g^{kl}\right.&&\nonumber\\
\left.-\partial^2_3g_{ij}+g^{kl}\partial_3 g_{kl}\partial_3 g_{ij}+2\partial_3\left(g_{ij}g^{kl}\partial_3g_{kl}\right)\right.&&\nonumber\\
\left.-\frac{1}{2}g_{ij}\left\{\partial_3g^{kl}\partial_3g_{kl}-\left(g^{kl}\partial_3g_{kl}\right)^2\right\}\right]&&\nonumber\\
-\frac{g^2_{_A}R^2}{2}\left({F_i}^kF_{jk}-\frac{1}{4}F_{kl}F^{kl}g_{ij}\right)+f_1\left(\hat{\Gamma}^3_{3k},\hat{\Gamma}^3_{33}\right)&=&0,\nonumber\\
\end{eqnarray}
where $f_1\left(\hat{\Gamma}^k_{k3},\hat{\Gamma}^3_{33}\right)$ refers to the terms which are proportional to $\hat{\Gamma}^3_{3k}$, $(\hat{\Gamma}^3_{3k})^2$, $\nabla_k(\hat{\Gamma}^3_{3l})$, and $\hat{\Gamma}^3_{33}$ which are zero with $\hat{\partial}_iR=0$, $\hat{\partial}_3R=0$, and $\partial_3A_i=0$.

Equations of motion for the 3D vector component $A_i$ are derived from the following variation
\begin{eqnarray}
\delta_{A}S_{\text{EH}}&=&\frac{M^2_{\text{Pl}}}{2}\int d^4x\sqrt{-g}\left[\delta_A\mathcal{R}-\frac{g^2_{_A}R^2}{4}\delta_A\left(F_{ij}F^{ij}\right)\right]\nonumber\\
&=&0
\end{eqnarray}
This variation leads to the equations of motion for $A_i$ as
\begin{equation}
\nabla_i F^{ij}+R^{-2}f_2\left(A_j,\hat{\partial}_jg_{kl},\hat{\partial}_3g_{kl}\right)=0.\label{Ai-equ}
\end{equation}
Note that, the second term in Eq. (\ref{Ai-equ}) is negligible because it is strongly suppressed by $R^{-2}$ which is in the order of $\Lambda_4$ ($\approx5.06\times10^{-84}$ GeV$^2$ \cite{Tanabashi2018}) as indicated in Appendix D. Thus, Eq. (\ref{Ai-equ}) can be approximated as $\nabla_i F^{ij}=0$.

Because the radion field $R$ would get an effective potential (generated from the cosmological constant term and one-loop quantum corrections) by which the size of the compactified dimension is physically fixed, we write the equation of motion for the radion field $R$ including this radion potential, corresponding to the variation of the 3D action (given in Appendix D) as follows
\begin{eqnarray}
\square R&=&\frac{1}{R}(\partial_iR)^2+\frac{R^2}{2M_3}\frac{\partial V}{\partial R}+f_3(A_i,F_{ij}),
\end{eqnarray}
where $f_3(A_i,F_{ij})$ is a functional of $A_i$ and $F_{ij}$ which vanishes for $A_i=0$.

\section*{Appendix C: A derivation of Eqs. (\ref{effEinsEq}) and (\ref{chiEq})}

In order to solve Eq. (\ref{3Dtensor-equ}), we use the method of the variable separation as $g_{ij}(x^i,x^3)=\chi(x^3)g^{(3)}_{ij}(x^i)$ where $g^{(3)}_{ij}(x^i)$ is the metric of the 3D effective theory whose wavefunction profile along the compactified dimension is $\chi(x^3)$. First, by substituting $g_{ij}(x^i,x^3)=\chi(x^3)g^{(3)}_{ij}(x^i)$ into Eq. (4), we find
\begin{eqnarray}
&&\mathcal{R}^{(3)}_{ij}-\frac{1}{2}g^{(3)}_{ij}\mathcal{R}^{(3)}+\Lambda_4\chi g^{(3)}_{ij}+\frac{1}{4R^2}\left[\left(\frac{1}{\chi}\right)^{''}\chi^2-\chi{''}\right.\nonumber\\
&&\left.+\frac{3\chi'^2}{\chi}+6\chi''-\frac{3\chi}{2}\left\{\left(\frac{1}{\chi}\right)^{'}\chi'-3\left(\frac{\chi'}{\chi}\right)^2\right\}\right]g^{(3)}_{ij},
\end{eqnarray}
where $\mathcal{R}^{(3)}_{ij}\equiv(\partial_k\Gamma^k_{ji}-\partial_j\Gamma^k_{ki}+\Gamma^l_{ji}\Gamma^k_{kl}-\Gamma^l_{ki}\Gamma^k_{jl})$ with $\Gamma^k_{ij}\equiv g^{(3)kl}(\partial_i g^{(3)}_{lj}+\partial_j g^{(3)}_{ki}-\partial_l g^{(3)}_{ij})/2$ and $\mathcal{R}^{(3)}\equiv g^{(3)ij}\mathcal{R}^{(3)}_{ij}$, and $'$ denotes the derivative with respect to the coordinate $x^3$. After simplification, we obtain
\begin{eqnarray}
\mathcal{R}^{(3)}_{ij}-\frac{1}{2}g^{(3)}_{ij}\mathcal{R}^{(3)}+\left[\frac{1}{R^2}\left(\chi''+\frac{11}{4}\frac{\chi'^2}{\chi}\right)+\Lambda_4\chi\right]g^{(3)}_{ij}=0.\nonumber\\
\end{eqnarray}
This equation can be written in the following way
\begin{eqnarray}
\frac{\mathcal{R}^{(3)}}{6}=\frac{1}{R^2}\left(\chi''+\frac{11}{4}\frac{\chi'^2}{\chi}\right)+\Lambda_4\chi.\label{VSE}
\end{eqnarray}
It is clear that the left-handed and right-handed sides of Eq. (\ref{VSE}) depend on the variables $x^i$ and $x^3$, respectively. This implies that these sides are equal to a constant denoted by $\Lambda_3$, which leads to
\begin{eqnarray}
\mathcal{R}^{(3)}=6\Lambda_3,\label{redEnEs}\\
\chi''+\frac{11}{4}\frac{\chi'^2}{\chi}+\kappa^2\chi&=&\frac{\Lambda_3}{R^{-2}},
\end{eqnarray}
$\kappa\equiv\sqrt{\Lambda_4}/R^{-1}$. Note that, Eq. (\ref{redEnEs}) is equivalent to Eq. (\ref{effEinsEq}) in the paper.

\section*{Appendix D: The radion potential}
The dimensional reduction of the 4D Einstein gravity on $S^1$ given in the Einstein frame reads
\begin{eqnarray}
S_{3D}&\supset&\int d^3x\sqrt{-g_3}\left[\frac{M_3}{2}\left\{\mathcal{R}^{(3)}-2\left(\frac{\partial_i R}{R}\right)^2\right\}\right.\nonumber\\
&&\left.-M_3\left(\frac{r}{R}\right)^2\Lambda_3\right],\label{4Deffact}
\end{eqnarray}
where $g_3\equiv\text{det}\left[g^{(3)}_{ij}\right]$ and the 3D Planck energy scale is identified as
\begin{eqnarray}
M_3&\equiv&rM^2_{\text{Pl}}\int^\pi_{-\pi}dx^3\chi^{1/2}\nonumber\\
&=&\frac{2\sqrt{\pi}\Gamma(17/30)}{\Gamma(16/15)}rM^2_{\text{Pl}}.
\end{eqnarray}
The last term in the action (\ref{4Deffact}) is the tree-level potential of the radion field generated by the dynamics of $3$D tensor component along the compactified dimension. In addition, the one-loop quantum corrections would contribute to the radion potential as 
\begin{eqnarray}
V_{\text{1L}}(R)&=&\sum_i(-1)^{s_i}n_iR\left(\frac{r}{R}\right)^3\rho_i(R)\int^\pi_{-\pi}dx^3\chi^{\frac{3}{2}}.\ \
\end{eqnarray}
Here $s_i$ is equal to $0(1)$ for the fermions(bosons), $n_i$ is the number of degrees of freedom corresponding to the $i$th particle, and the Casimir energy density with respect to the $i$th particle is given by \cite{Arkani-Hamed2007}
\begin{eqnarray}
\rho_i(R)=\sum^{\infty}_{n=1}\frac{2m^4_i}{(2\pi)^2}\frac{K_2(2\pi nm_iR)}{(2\pi nm_iR)^2},
\end{eqnarray}
where $m_i$ and $K_2(z)$ are the mass of the $i$th particle and the modified Bessel function, respectively. It should be noted here that due to the function $K_2(z)$ suppressed for $z\ll1$ the particles with their mass which is much larger than $R^{-1}$ do not contribute significantly to the one-loop term of the radion potential and hence we can ignore their contribution. On the other hand, only the light degrees of freedom contribute significantly to $V_{\text{1L}}(R)$. 

\begin{figure}[h]
 \centering
\begin{tabular}{cc}
\includegraphics[width=0.4 \textwidth]{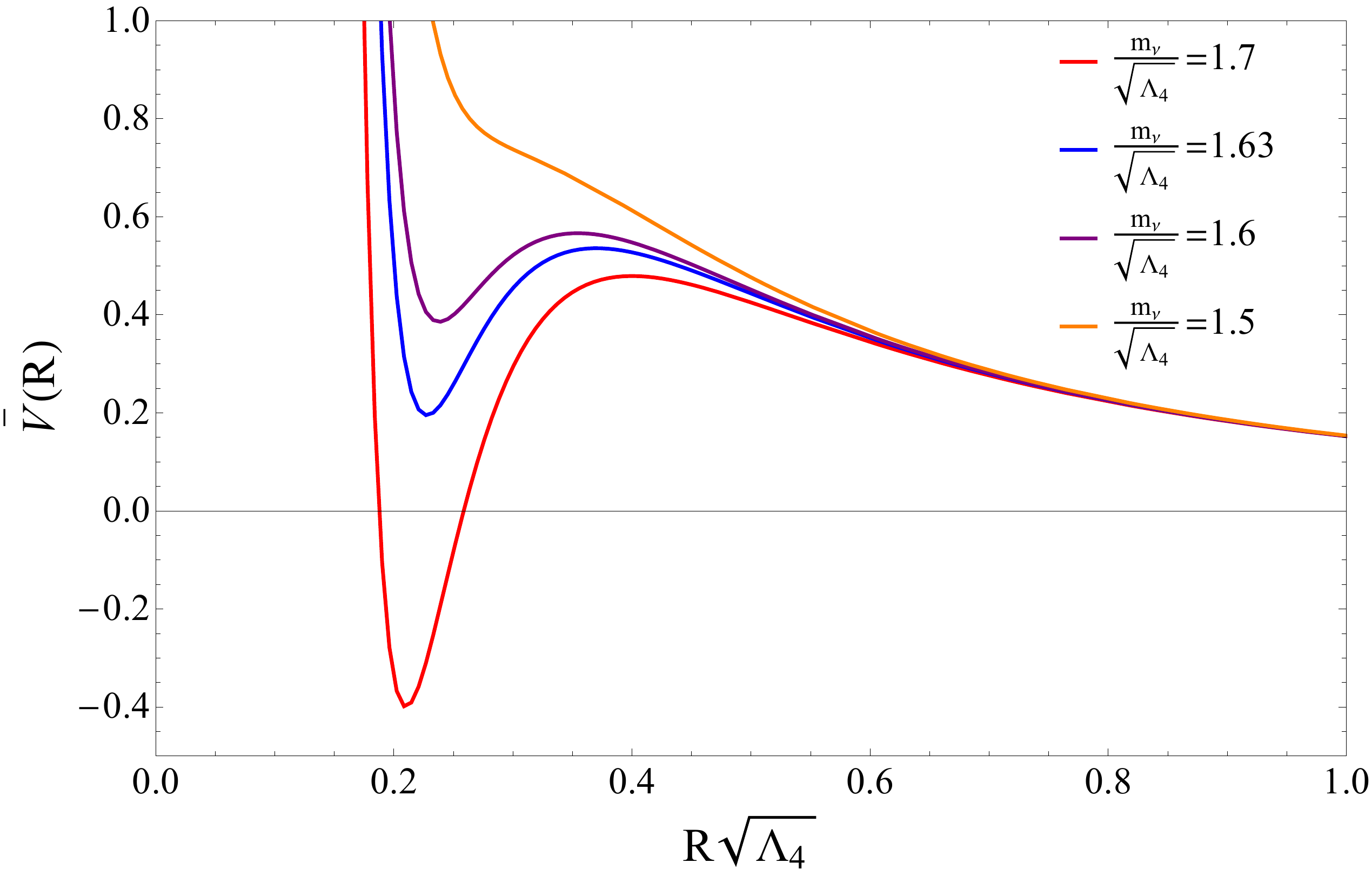}
\end{tabular}
 \caption{The scaled radion potential $\bar{V}(R)\equiv V(R)/(2\sqrt{\pi}r^3\Lambda^3_4)$ versus the scaled $S^1$ radius $R\sqrt{\Lambda_4}$ for various values of $m_\nu/\sqrt{\Lambda_4}$ with $M^2_{\text{Pl}}\Lambda_3/\Lambda^2_4=0.1$. The red and orange curves correspond to the AdS vacuum and no vacuum, respectively, whereas the blue and purple curves lead to the dS vacuum.}\label{radpot}
\end{figure}

The radion potential $V(R)$ thus is a sum of the tree and loop level contributions, which is expanded in terms of $m_iR$ for $m_iR\ll1$ as
\begin{eqnarray}
\frac{V(R)}{2\sqrt{\pi}r^3}&\simeq&\left[\frac{\Gamma(17/30)}{\Gamma(16/15)}\frac{M^2_P\Lambda_3}{R^2}+\frac{1}{16\pi^2}\frac{\Gamma(7/10)}{\Gamma(6/5)}\frac{1}{R^6}\right.\nonumber\\
&&\left.\times\sum_i(-1)^{s_i}n_i\left\{\frac{1}{90}-\frac{(m_iR)^2}{6}+\frac{(m_iR)^4}{48}\right\}\right].\nonumber\\
\end{eqnarray}
The behavior of the radion potential is depicted in Fig. \ref{radpot}. Here, the bosonic contributions come from graviton and photon, whereas the fermionic contributions come from the lightest neutrino and an additional massless Dirac fermion. 

\end{document}